\documentclass[journal,10pt, a4paper,onecolumn,final]{IEEEtran}

\usepackage{subcaption}
\usepackage{optidef}
\usepackage{xcolor}
\usepackage{ulem}
\usepackage{lipsum}

\usepackage{tabularx}
\usepackage{multirow}
\usepackage{booktabs}
\usepackage{diagbox}

\newcommand{\superscript}[1]{^{\text{#1}}}

\def \highlightcolor {black}

\newcommand{\revise}[2]{#1}
\newcommand{\secrevise}[2]{#1}
\newcommand{\thirdrevise}[2]{#1}

\begin{document}

\title{5G Network Slice Isolation}


 \author{Stan Wong, Bin Han, and Hans D. Schotten
	 	\thanks{Stan Wong is with the Hong Kong Telecommunication Limited, Email: stan.ws.wong@pccw.com. B. Han and H. D. Schotten are with the University of Kaiserslautern, Emails: \{binhan, schotten\}@eit.uni-kl.de. Hans D. Schotten is with the German Research Center of Artificial Intelligence (DFKI GmbH), Email: hans\_dieter.schotten@dfki.de.}
 }


\maketitle

\begin{abstract}
This article reveals an adequate comprehension of basic defense\revise{, security challenges,}{} and attack vectors in \revise{deploying}{} multi-network slicing. Network slicing is a revolutionary concept of providing mobile network on-demand and expanding mobile networking business and services to a new era. The new business paradigm and service opportunities are encouraging vertical industries to join and develop their own mobile network capabilities for enhanced performances that are coherent with their applications. However, a number of security concerns are also raised in this new era. In this article, we focus on the deployment of multi-network slicing with multi-tenancy. We identify the security concerns, and discuss about the defense approaches such as network slice isolation and insulation \revise{in a multi-layer network slicing security model. Also, we identify the importance to appropriately select the network slice isolation points, and propose a generic framework to optimize the isolation policy regarding the implementation cost while guaranteeing the security and performance requirements}{}.
\end{abstract}

\begin{IEEEkeywords}
5G; network slicing; security; isolation; insulation
\end{IEEEkeywords}

\IEEEpeerreviewmaketitle

\section{Introduction}
\label{intro}
Network slicing \revise{\cite{EURSIP:NS5G}}{} is a revolutionary concept of enabling mobile network on-demand. It extends the business model of the mobile networking from the traditional tariff subscription to the new cloud computing paradigm: network slice as a services (NSaaS). \revise{The basic principle of network security such as authentication, authorization, confidentiality, integrity and availability can be found in \cite{Wiley:NSSecC}.}{} The new business model and service opportunities are motivating vertical industries to join and develop their own mobile networks, and specify the network infrastructure capabilities and performance to align with their business and application characteristics. To achieve this aim, it requires to define an adequate comprehension of the defense mechanisms that protect all deployed network slices of various types with different network performance and security requirements \revise{\cite{IEEE:5GNSSec}}{}. Especially, these defending mechanisms have be considered not only for the traditional physical network infrastructures, but also in a nested virtualized network environment.

Formerly, the traditional network infrastructure had been considered as a secure environment because it was operated under a single administrative domain and fully maintained by an operator. However, this has been changed in NSaaS, where the mobile network operator (MNO) or network slicing service provider offer network slices of various types for leash, which coexist over a shared physical infrastructure \revise{\cite{MDPI:NSC}. Jos\'e Mar\'ia Jorquera Valero et al. \cite{JNSM:5GMD} pointed out that multi-tenancy requires a multi-domain security and risk management embedded into the design. Also, the current trust models and exiting defense methodologies might not be appropriate to the operation of NSaaS.}{.} \revise{Those}{Such} network slices may have different levels of security demand and various specifications of tailored network protection measures. In such an emerging network environment, the defending and attacking surfaces become much wider than those in legacy systems, and the security problem becomes more complex than the mobile industry had anticipated. Therefore, it calls for an effective isolation mechanism and policy approaches that can protect the network from attacks and infections over these new wide surfaces. \revise{For example, Tomasz Wichary et al.\cite{MDPI:NSSecCrlt} discussed to security control and policies to protect those network perimeters and Chun-I Fan et al. \cite{IEEE:CAuC5GNS} developed a scheme for cross-network slice authentication which against the attackers who aim to impersonate users, network operators or network slices by providing a secure session key exchange.}{}

For developing security solutions to protect the complex network environment when deploying NSaaS, it is seriously important to identify the possible attacking vectors, the adequate defending mechanisms, and the appropriate security technologies. This identification process also helps us to gain practical security knowledge and increase the security awareness. 

Basically, NSaaS shall provide various levels of isolation, e.g. application segmentation isolation, virtual machine isolation, network segmentation isolation, and resource isolation, to secure the infrastructure of mobile network operator (MNO) and to protect the tenant's privacy as well as their services. On the other hand, a typical network slice tenant generally expects its network slice to run as a standalone and fully independent mobile network. Neither such a network slice nor the data of its tenant shall be accessed by other unauthorized tenants. Unfortunately, network slice is based on virtualization, containerization and software-defined network technologies, faults and mistakes can propagated to other network slices via the virtualized environment and attackers may cross network slices to misuse the networks for their purposes. This is the main reason network slice isolation that becomes the primary goal of the mobile industry to resolve in order to deploy a secure NSaaS. \secrevise{In their survey in 2018~\cite{CESAR:NSSecurityAI}, Luis Su\'arez et al. have identified network slice isolation as the core concept that impacts the network slice security. Since the traditional approaches such as traffic isolation, encryption, and firewalls are not providing sufficiently satisfactory performance in countering the related cyber-threats, they have envisioned some possible artificial intelligence mechanisms to enhance the network slice security. Unfortunately, little progress has been reported in the suggested approaches. One of the main reasons for this lack of achievement can be the absence of deep and thorough understanding of the security model in slice isolation.}{}

This article is to demystify the appropriate defense mechanisms and to provide adequate isolation approach in different point of network slice. This isolation point must be selected based on the characteristics of the network slice and the MNO's network infrastructure strategy. \revise{The main novel contributions of our work, in this article we have: 1) identified security challenges in deploying NSaaS, 2) proposed a multi-layer model to decompose the network slicing security complexity, 3) analyzed the impact of network slice isolation point selection, and 4) proposed a framework to optimize the selection of network slice isolation points.}{}

This article is organized as follows: Section \ref{sec2} provides the principles of network slicing and network slice types of characteristics. Section \ref{sec3} discusses the challenges of network slicing security when deploying a NSaaS platform. We use a multi-layer approach to explain the complexity layer-by-layer, then identify precision of network slice isolation would affect the defense and performance of the network slice. Subsequently, we develop a mathematical model of network slice isolation in relating to the level of control MNO and tenant that would apply to the cost of deployment network slice relationships in Section \ref{sec4}. Finally, we conclude the paper in Section \ref{con}. 

\section{The Principles of Network Slicing and Network Slice Types}
\label{sec2}
Network slicing is a logical network representation, composed with specific mobile network infrastructure configuration, which consists of various levels and types of isolation in a physical infrastructure. It is basically enabled by virtualization, containerization, software-defined network (SDN), virtual network function (VNF) service chain, network function virtualization (NFV) \revise{\cite{Wiley:SS5GNS}}{} and flexible transport network technologies. MNO's is expected to utilize those technologies to provide a secure network environment across radio access network, transport network and core network. This secure network environment shall be fully optimized with multiple network slice coexistence and their different service characteristics and requirements. On the other hand, tenant is expected their network slices structure as a standalone and fully independent mobile network. Besides, other tenants shall not have an unauthorized access to their network slices nor unauthorized interception with the other tenants' data.

Network slicing dynamically gives MNO a flexibility in organising, coordinating and orchestrating any available resource in the wireless and wired network environment. Those resources can be differentiated into a specific service in a particular location. For example, a manufacturer customer would like to have a network slice with a particular location within few cell sites only. A utility company would like to have a smart grip network slice in some remote sites. Another case, hospital authority customer would like to have a network slice within a hospital area. Those three typical cases illustrate network slice services that can be dynamically deployed and provision in a unique geolocation. Also, these individual network slices can port to other network slice service provider or MNO network slice platforms. GSM Association (GSMA) has provided an introduction of network slice \cite{GSMA:NSintro}, and has been proposed the Network Slice Generic Template for formulate menu for selecting the network slice perimeters. This GST model can be converted to network provisioning data model for deploying network slice purpose \cite{GSMA:NSGT}.

\section{Network Slice Isolation as a Security Measure}
\label{sec3}
NSaaS is set to deliver an on-demand mobile network. It encourages vertical industry to design and develop their mobile network infrastructure and mobile network service. These mobile network infrastructures and services utilize virtualization, containerization and SDN technologies to increase the flexibility of network provision, deployment and operational models, and the business transformation and service agility across multiple mobile networks. Particularly, these mobile network infrastructures or services provides network independence and network seclusion\revise{, which has been demonstrated with multiple points-of-presence slice segment stitching to construct a network slice and also various resources  being flexibly manipulated for a network slice \cite{IEEE:CloudNet}.}{.}
Traditionally, MNO only has a single administrative domain (AD) to manage, an network element and subscriber to protect, an impersonation of subscriber to prevent, and a static attack vectors to identify etc. However, when NSaaS is \revise{}{being} deployed, \revise{}{due to} the network flexibility and service agility \revise{will}{that} lead to a number of new security challenges. \revise{In this paper, we provide a comprehensive study on security challenge in four aspects from identify the protection assets, prevent attacks and human errors, identify the right selection of isolation points and different assets require to manage. Particularly, }{}for ensuring the understanding of NSaaS new security challenges and applying the right NSaaS operation protection without affecting the network slice service performance requirement that is vitally important in an multi-network slicing environment. It is also critical that the NSaaS security perimeters are adequately defined throughout the entire NSaaS security chain and in operational level from radio access network to transport network, and from transport network to core network. 

\subsection{Challenges in Network Slicing Security}
In this subsection, the key network slice security challenges are defined in four aspects which are protection, prevention, identification and management\revise{, as summarized in Table \ref{tab:challenges}}{}.

\begin{table}[!hbtp]
	\centering
	\caption{Identified challenges in network slicing security}
	\label{tab:challenges}
	\begin{tabular}{lm{5cm}m{5cm}}
		\toprule[2px]
		\textbf{Aspect}	&\textbf{Subject}								&\textbf{Objective}\\
		\midrule[1.5px]
		Protection		&network infrastructure							&network resilience and service availability\\
		&&\\
		Prevention		&unauthorized access and inappropriate use		&cross-AD resource isolation and robustness to insider threat\\
		&&\\
		Identification	&security threats								&establishing appropriate security control policies\\
		&&\\
		Management		&ADs, virtual environment visibility, subscribers of tenants	&increased virtual environment visibility and reduced network risk\\
		\bottomrule[2px]
	\end{tabular}
\end{table}

The protection challenges \revise{are raised by concerns about}{of} network infrastructure \revise{to}{} support NSaaS, \revise{where it}{that} shall begin to consider the protection of network infrastructure from static resources to dynamic resources network environments. Typically, static resources can be referred to hardware assets and dynamic resources can be considered as software assets. Furthermore, these software assets can be created at runtime when network elasticity is triggered by traffic and network services on-demand. Since, these runtime software assets can be network slices, virtual network functions and SDN properties that may overload the network and affect the network services availability. Therefore, we have to protect the network availability, service reliability\revise{}{'s} and company liabilities at all time. Particularly, other network services are having functional error or being compromised which can be possible to affect any other network services availability. All these protections shall be considered from network resilience to risk assessment of network services.

The prevention challenges are the unauthorized access and inappropriate use of network infrastructure resources, which can be considered the access or usage from the same AD or from another ADs. Traditionally, MNO only manages a single AD and never \revise{has}{have} experience on managing and authorizing 3rd parties that access to various level of resources based on the services level agreement with the tenant. Therefore, preventing cross ADs resource access is another challenge MNO requires to manage. Particularly, under the virtualized network environment, co-resident attacks may trigger an unauthorized access to another virtual machine co-existing under the same bear-metal. Furthermore, MNO also requires to prevent another serious issue in all kinds of system within the infrastructure which is insider threats. In order to prevent insider threat under such fast evolve and change network environment, a proper management process or control has to apply on top of traditional approach\revise{es}{}, for example, ISO/IEC 27001 has a series control processes to ensure the information security management in securing the system. We often face an unknown \revise{threat}{thread} when network automation applying to virtualized network infrastructure environment, there is possibility that attacker may be inappropriately manipulated network resources via auto-optimization and auto re-configuration. Therefore, we shall apply zero trust to prevent auto-manipulation of network resources. 

The identification of security threats challenges \revise{is typically an essential task for MNO}{are typical MNO that requires to do} before the network deployment. Usually, MNO will establish security control policies appropriately which is not just based on the local regulations requirements and international benchmark approaches \cite{CIS:url}, \revise{but}{and} also \revise{demanding}{requires} to adapt the best practice from the industry. Therefore, identifying the security control policies for deploying NSaaS requires to consider the security policies under the flexible network and dynamic network runtime environments, it cannot simply apply black-box approach\revise{es}{} that will eventually expose various unidentified attack vectors and vulnerable loopholes. Since, the common practice of identifying the attack vectors or conducting the risk assessments requires an existing network environment. Particularly, attack vectors will not be straight forward without an existing network infrastructure and services environments. Even though, the flexible network infrastructure is unpredictable on the managing the resources but we shall clearly state out the security policies when applying network elasticity. Furthermore, we also have to identify \revise{}{each of the network slice} the adequate physical and logic isolation points \revise{for each of the network slices}{} to protect the service availability, set the security perimeters\revise{,}{} and provide appropriate security measures in the future. 

In terms of the network management challenges, we have a number of items that \revise{must}{required to} be seriously considered. The MNO shall provide policies to manage the unknown ADs and the virtual environment visibility. Especially, virtual environment visibility can be managed by different technological techniques e.g., micro segmentation, hypervisor firewall etc. Those techniques can increase the visibility, but also \revise{require}{required to have} a substantial knowledge to manage them. On the other hand\revise{}{s}, under NSaaS, we have many tenants that need\revise{}{ed} to manage. For example, tenant's identity, access and privacy shall be properly managed. Also, the MNO shall provide a privacy scheme or guideline for tenants to manage their subscribers in order to reduce the risk of the network. 

The above four aspects can assist MNO to achieve a secure operation of NSaaS, we propose to plan and provide a precise policy of control in fulfilling \revise{them}{the above list} as the basic requirements.

\subsection{Decomposition of Network Slicing Security Complexity}
\label{sec32}
\begin{figure}[!t]
	\centering
	\includegraphics[width=5.5in]{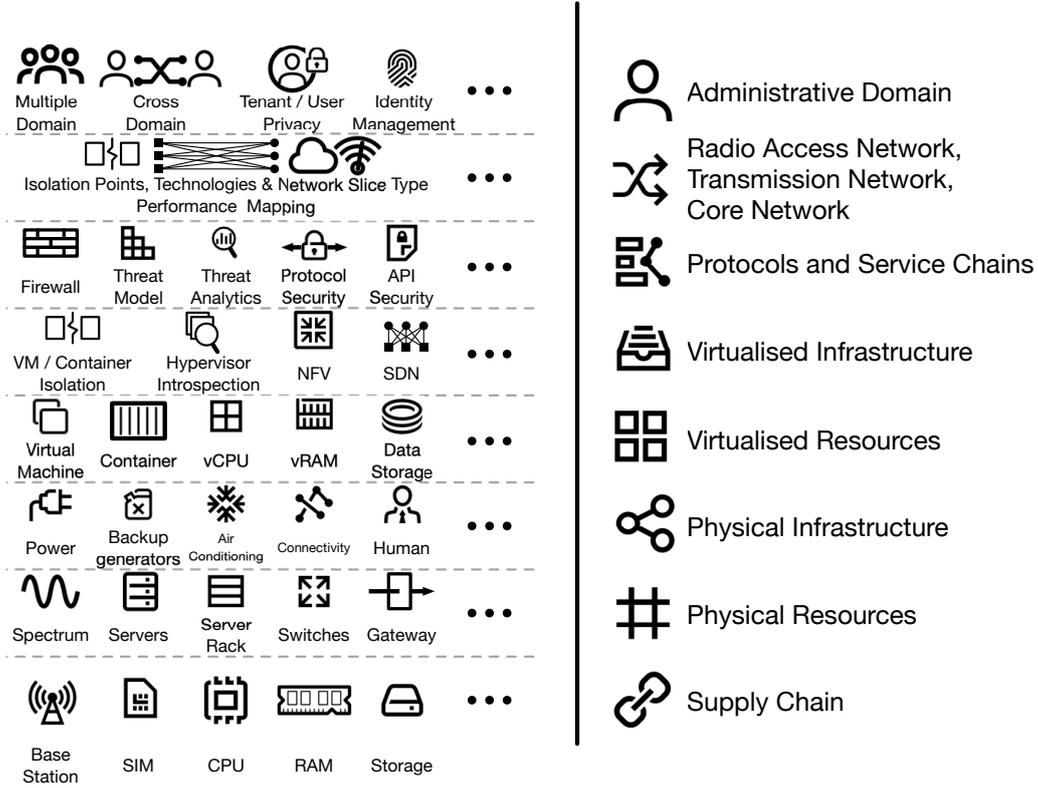}
	\caption{Network slice basic elements}
	\label{fig:Netslice_ele}
\end{figure}

In this subsection, \revise{we present a multi-level model of the network slicing security decomposition}{a decomposition of the network slicing security complexity is presented in Figure \ref{fig:Netslice_ele}. Figure \ref{fig:Netslice_ele} is illustrated a multi-level model of the network slicing security decomposition}. \revise{Basically, this model also represents a network construction sequence which starts from deciding the type of devices available in the supply chain. Once installed to the network, those devices become physical resources that formulate an infrastructure. Supposed to be fully utilized, they can be transferred into virtual resources by applying virtualization and containerization technologies. Consequently, those formulated virtual resources should be managed by an information management platform e.g., NFV. After the physical and virtual infrastructures being fully established, we start to consider the protocol and service chains protection methodologies and the appropriate isolation points in the network slice. Finally, from the MNO's point of view, it is essential to consider a network slice platform to manage the network slice tenants by means of tenant identities, access rights, services, etc. Note that the above description is simplified regarding the deployment consideration and sequence of architecture design decisions. Also, between each two layers, there is a tight relationship and logical linkage in the deployment of a network slice. Furthermore, each of the layers and elements has a specific protection method, which we are going to discuss in this subsection.}{} 

\revise{As illustrated in Figure \ref{fig:Netslice_ele}, the}{The} lowest three levels \revise{in our model are}{} are \revise{inherited from}{} the traditional network security \revise{model,}{} which \revise{concern}{are} the fundamental of telecommunication equipment supply chain security, physical resource security \revise{}{concern} and physical infrastructure\revise{, respectively.}{ concern.} From the fourth \revise{}{layer} to the top layer are the logical and information security concern\revise{s}{} which are considered \revise{to deal with a}{} wider attack surface \revise{by every next layer}{layer by layer}. Furthermore, the complexity of defense in each layer will also increase layer-by-layer from the bottom to the top layer. We further describe each of layer characteristics in the following; 

\emph{Layer I Supply Chain} -- Usually, it is a first line of defense and is not only considered a physical active electronic component and passive electronic component. Software components or entities shall be included within the supply chain. Those \revise{components'}{component's} software is often employed \revise{with}{} malicious code. Therefore, we have to have certain level of control \revise{over}{of} supply chain when deploying NSaaS. ISO 28000 specification has a well-established supply chain security management control framework that can be applied. NIST has also suggested a supply chain life-cycle management \cite{NIST:SC}. Furthermore, supply chain security management is not just to deliver control and assurance to the overall system, it also requires to define the level of control processes, certifications of the product within the best practice in the current time, and the trustworthiness of the protocol applied to test the products. GSMA provides a supply chain toolbox to give a guideline of this first line of defense \cite{GSMA:SC}, and NSCS also provides 12 principles to ensure the first line of defense under control within the appropriate stage of the overall supply chain \cite{NCSC:urlSC}.

\emph{Layer II Physical Resources} -- Often, MNO unifies the physical network elements and license's components as physical resources which will increase the flexibility of overall mobile network infrastructure, and refine the productivity by applying different service management methodologies. Furthermore, MNO also constantly searches various methods and techniques to fully utilize all available resources in their network infrastructure. Also, deploying network slice, the second line of defense is to manage different types of physical resources that apply to a particular network slice. For example, a critical infrastructure network slice can be only deployed in a few specific locations with selected spectrum thereat, and the local breakout may also require to be deployed with an air-gap isolated server rack, switch and the internet gateway.  

\emph{Layer III Physical Infrastructure} -- Facility infrastructure resiliency gives service reliability to the MNO's mobile network infrastructure. There are number of international data \revise{center}{centre}control frameworks \cite{ISO:22237, EU:50600} to protect this third line of network slice defense’s service availability and reliability. For example, an utility smart grid network slice may request\revise{}{s} a wide area deployment and require\revise{}{s} a certain level of service availability and reliability. Hence, MNO may need to pick the right level of data \revise{center}{centre}for such network slice deployment. Often, mobile network infrastructure is constructed by different data \revise{centers, which }{centre. We often employ} different data \revise{center}{centre}management teams and companies \revise{are often employed}{} to manage\revise{}{ the data centre}. In maintaining the data \revise{center}{centre}service reliability and ensuring the different level of data \revise{center}{centre}security, data \revise{center}{centre}security is not only facility security and also include identity and access management etc.  

\emph{Layer IV Virtual Resources} -- Generally, network slicing is based on virtualization and containerization technologies as its foundation. Network slices can be constructed under virtual machines, containers or combination of virtual machines and containers, and each of the network slice can be specifically restricted on the number vCPU, vRAM and the type of storage. The MNO requires to manage its virtual resources, so that it does not exceed the maximum level of physical resource limitation and cause service interruptions. 

\emph{Layer V Virtual Infrastructure} -- The level of complexity in this layer has been significantly increased. We have to consider the implementation virtual machine and container isolation techniques to avoid co-residency attack. The typical technique would apply is the hypervisor introspections or serverless container isolation technique in kernel level. The virtualize infrastructure can have access control list for particular application to secure the entire network segment using micro-segmentation which automatically apply network segregation. Therefore, the virtualization and containerization network security would be the main consideration in this layer. Since, this layer’s defense is across different area\revise{s}{} of technology implementation\revise{,}{} from application to virtual network \revise{segmentation,}{segment} and from infrastructure access control to CPU firmware trust model. All these techniques are trying to keep network slice isolate\revise{d}{} from each other.

\emph{Layer VI Protocol and Service Chain} -- In this layer, a formulated network slice shall have a specific service to deliver. Usually, MNO formulates those services that may use service chain approach. This service chains often are in a sequential manner of network functions which function can also split into multi-locations and the traffic would propagate from one network location to another in a specific sequence. Due to network service chain sequential structure, we can collect network intelligence data which can be used to increase the virtual network infrastructure visibility and threat intelligence protection on different level of network slice defense. On the other hand, we have to avoid the inappropriate of virtual resource manipulations, therefore, we can use appropriate security protocol and the API security to prevent malicious manipulations.

\emph{Layer VII Radio Access Network, Transmission Network and Core Network} -- When deploying network slice, we need to identify various isolation points as network defense perimeters\revise{, where different isolation techniques can be applied to}{. There are different isolation techniques can be applied to the isolation point}. Those isolation points must be carefully selected, otherwise, the service performance can be easily affected. Therefore, mapping the isolation points with adequate technology under different network slice type is an important \revise{process in}{processes} deploying network slices.

\emph{Layer VIII Administrative Domain} -- Consequently, there is a possibility, tenant may have purchase multiple network slices across different MNOs, and tenant may share all resources across multiple network slices. Therefore, MNO or network slice service provider requires to protect each AD's user and tenant privacy, and must manage users’ and tenants’ identities who accesses the appropriate AD. 

The above multi-layers approach can assist network slice service provider or MNO to distinguish and differentiate the level of managing a NSaaS platform and to protect the overall MNO network service availability. After resolving the network slice complexity in layers, we shall focus on the practical deployment of NSaaS which would focus on the defense of 3 domains in data centre; radio access network, transport network and core network. 

\subsection{Precision of Network Slice Isolation Point}
\label{sec33}
Identifying an adequate network slice isolation point and applying right network slice isolation mechanism and policy at those isolation points are the main challenges in deploying multi-network slicing to a mobile network operator network. Network slice\secrevise{s are}{} designed to \secrevise{support the co-existence of multiple tenants}{be supported multi-tenant co-exist} on an MNO physical network with independent, isolated and fully secured network service\secrevise{s}{}. Also, \secrevise{one}{} tenant would not know another tenants' existence in the network. \secrevise{A similar strategy has been proposed on the Internet to isolate services or applications using service oriented architecture \cite{IEEE:IsoApp}. However, it might need abnormal detection to protect the behavior of the network slice from faults, e.g., an inappropriate selection of isolation points. In case of such faults, the anomaly detection algorithm can as well be invoked to obtain the score of isolation points behavior\cite{IEEE:know}, which may be further exploited by machine learning techniques to isolate the faults \cite{NASA:ML} and to model the slice behavioral patterns under a particular setup of isolation points. }{}

GSMA has defined 8 types of network slice use cases and each of the network slice type could have different network configurations, network performance requirements, traffic criteria and security control etc. All these characteristics would ultimately lead to deliver service experience to the subscriber and fulfil the network slice Service Level Agreement (SLA) \revise{under a secured \cite{Wiley:e2eNS}}{}. Especially, multi-network slice deployment involves with different network technologies, resource migration and resource optimization at the runtime. \revise{Either an inappropriate selection of the isolation points or wrongly}{Wrongly} applying isolation mechanism and policy in each of the isolation point can cause a network performance degradation or service delivery interruption after resource optimization and migration. Therefore, we \revise{shall}{require to} identify each of the possible isolation point and adequate security mechanism and policy applying to those isolation points. \revise{By appropriately specifying these features it helps not only by securing the network slice, but also by enhancing the network performance without affecting the subscriber experience or violating the SLA.}{No just can secure the network slice and also can provide network performance without affecting the subscriber experience, and violate the SLA.}

\begin{figure}[!t]
	\centering\includegraphics[width=.9\linewidth]{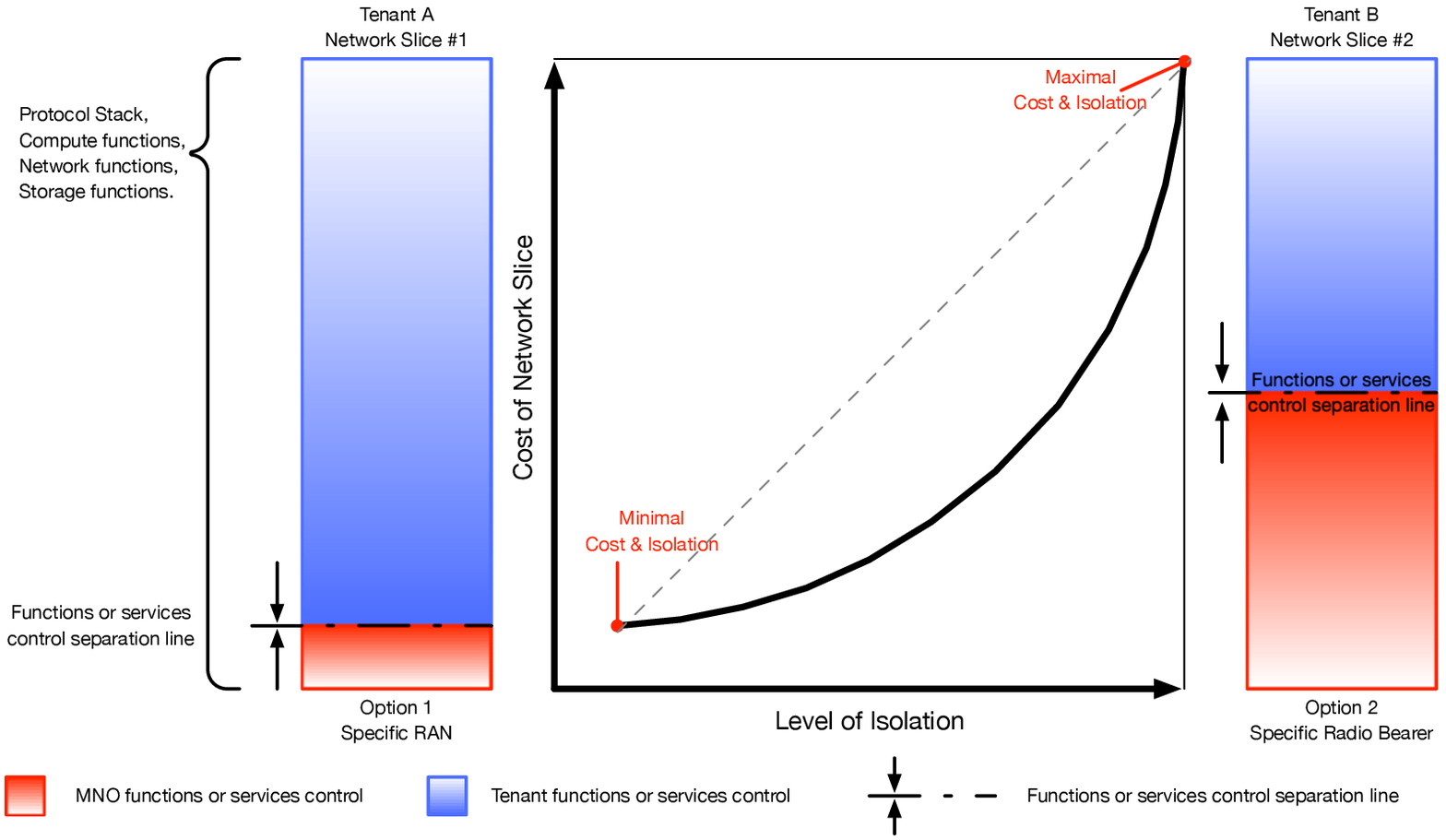}
	\caption{\thirdrevise{The network slice cost raises along with the isolation level due to extra implementation of tenant-dedicated functions and service control}{The level of isolation of network slice in relating to protocol stack and service}.}
	\label{nsi1}
\end{figure}

\begin{figure}[!t]
	\centering\includegraphics[width=3.55in]{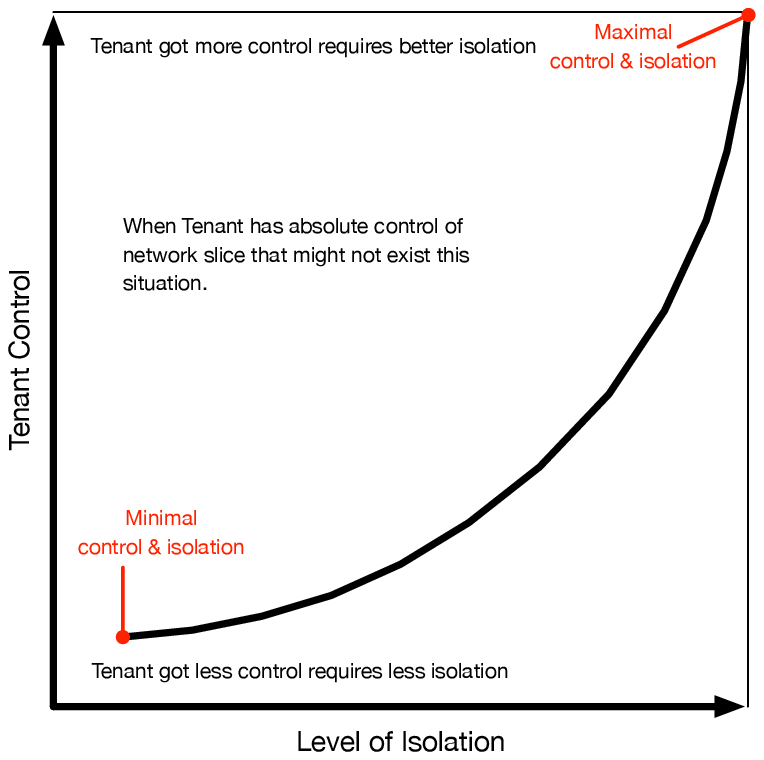}
	\caption{\thirdrevise{The tenant obtains more control on its slice with a higher isolation level, since more shared functions are replaced by dedicated ones}{The level of isolation from tenant control and in relating to tenant network slice deployment method selections}.}
	\label{nsi2}
\end{figure}

\begin{figure}[!t]
	\centering\includegraphics[width=3.55in]{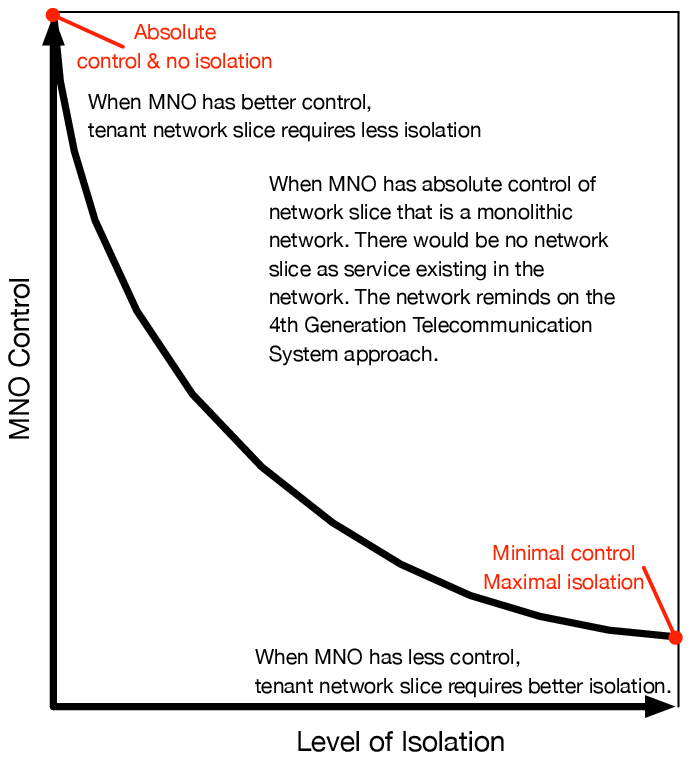}
	\caption{\thirdrevise{The MNO retains less control on a leased slice with a higher isolation level, since more shared functions are replaced by dedicated ones}{The level of isolation from MNO control and in relating to MNO network slice deployment method selections}.}
	\label{nsi3}
\end{figure}

\begin{figure}[!t]
	\centering\includegraphics[width=3.55in]{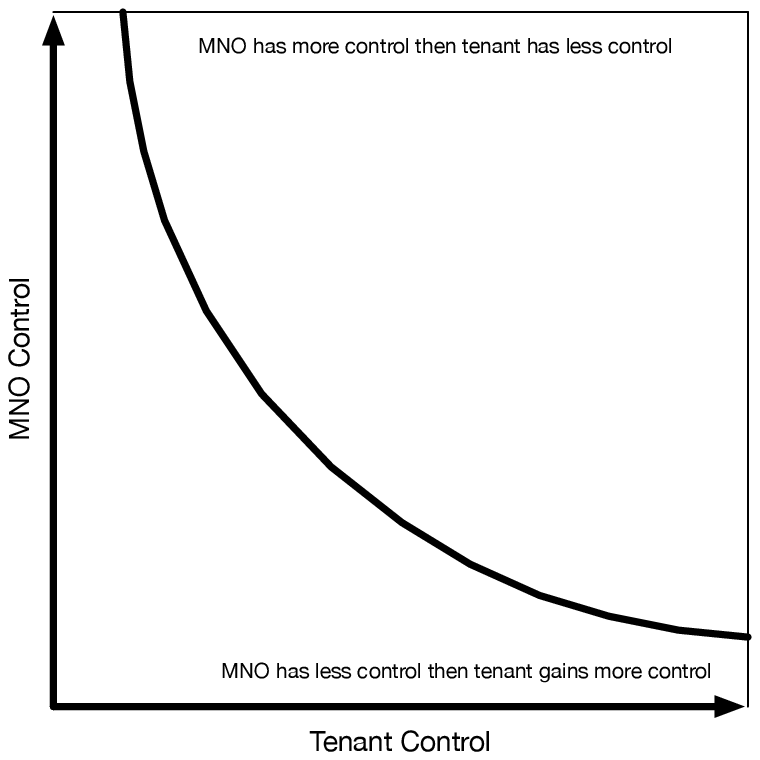}
	\caption{\thirdrevise{The controls of MNO and tenant are exclusive to each other}{The level of MNO control in relating to Tenant control in deploying of network slice}.}
	\label{nsi4}
\end{figure}

Figures~\ref{nsi1}--\ref{nsi4} provide \revise{illustrations of some phenomena}{an illustration of the phenomenon} when deploying network slice. Figure~\ref{nsi1} \revise{is}{has} divided into three parts; on the right and \revise{left sides}{left-sides}, \revise{two options are illustrated where the tenant requests for a network slice with the most tenant control and minimal influence from MNO (left), or balanced control shared between the tenant and the MNO (right), respectively. In the earlier case,}{tenants have requested a network slice that is almost controlled entire protocol stack and network functions. It also shows that has very limited influence from MNO. In this case,} MNO only provides physical resources (e.g., spectrum etc.)\revise{; in the latter case, several layers of the protocol stack and some specific network functions are defined and controlled by the MNO}{}. In the middle, \revise{Figure}{figure}~\ref{nsi1} \revise{shows how the level of isolation matters to the cost of deployment when considering network slice isolation}{is shown when deploying network slice requires to consider the level of isolation that shall be applied and the cost of deployment network slice isolation in relating to each level of isolation}. Particularly, \revise{Figure}{figure}~\ref{nsi1} indicates the minimal and maximal cost of isolation that would start on a positive manner due to the physical resources (e.g., spectrum etc.) that \revise{}{are} belong to MNO. The graph also indicates the characteristics of the isolation relationship in between the level of control tenant that can be gained when deciding to purchase a network slice. Also, the graph indicates that is not directly proportional to each other due to the vast amount of isolation technique\revise{s}{} that can be applied to deliver similar protections. Figure~\ref{nsi2} provides an overview of controlling of a network slice by tenant. When tenant has a minimal control of network slice that implies the tenant fully \revise{relies}{rely} on MNO to manage the network slice, and the MNO has less responsibility to put isolation to the protect the network slice. On the other hand, when tenant has a maximal control of the network slice and the MNO is responsible to put isolation into the network slice for protecting the other tenant privacy. Figure~\ref{nsi3} reflects the control of MNO which is correlated to \revise{Figure}{figure}~\ref{nsi2}. Furthermore, \revise{Figure}{figure}~\ref{nsi3} indicates when MNO has absolute control of network slice that is a monolithic network. There would be no NSaaS existing in the network. The network reminds on the 4th generation telecommunications system. Finally, \revise{Figure}{figure}~\ref{nsi4} \revise{shows the exclusive relation between the MNO control and tenant control on any certain network slice}{is shown the relationship and characteristics the level of control MNO that shall have in relating to tenant controls of the network slice}. \revise{It shall be noted that Figures~\ref{nsi1}--\ref{nsi4} show no quantitative results but only qualitative relations among the level of isolation, slicing cost, and control levels, which can be straightforwardly derived from the control sharing mechanism and the cost budget of network slice isolation.}{}

\section{Model of Network Slice Isolation}
\label{sec4}
In this section, we develop a mathematical model to help MNO in identify the cost of deploying right network slice isolation points. We use 3GPP protocol stack \cite{3GPP:TS38300, 3GPP:TS33501} as a network slice logical deployment representation. We begin with the case where the isolation points can be independently selected for every individual slice. Ideally, those isolation points would not have any impact to the performance nor implementation cost of other slices. Therefore, we define a mathematical model to provide a rational representation. We start from the following notations:

We consider the set of all $N$ slices $\mathcal{N}=\{1,2,3\dots,N\}$, where every slice needs to implement a full stack $\mathcal{P}\revise{=\{1,2,3\dots,P\}}{}$ of protocol layers. Every individual slice $n\in\mathcal{N}$ can independently and flexibly choose the method to implement each protocol layer, either in a physical way or in a virtual way. This can be formulated with a binary indicator for every pair of network slice $n$ and protocol layer $p$:
\begin{equation}
	v_{n,p}=\begin{cases}
		1&\text{$n$ virtualizes $p$}\\
		0&\text{otherwise}
	\end{cases},\quad\forall n\in\mathcal{N}, p\in\mathcal{P}.
\end{equation}

For each protocol layer $p$ of a specific network slice $n$, call its isolation level $i_{n,p}$, the tenant control level \revise{$t_{n,p}$ can be selected from a finite discrete set $\mathcal{T}_{n,p}\subseteq\mathcal{T}$, where $\min(\mathcal{T}_{n,p})=0$ and $\max(\mathcal{T}_{n,p})=t_{n,p}\superscript{max}\in(0,1)$. We also define the MNO control level $m_{n,p}=1-t_{n,p}$,}{$t_{n,p}\in{[0,t_{n,p}\superscript{max}]}$ where $t_{n,p}\superscript{max}\in(0,1)$, the MNO control level $m_{n,p}\in(0,1]$,} the operations cost $c_{n,p}\superscript{op}$, and the infrastructural cost 
\begin{equation}
	c_{n,p}\superscript{ifr}=c_{n,p}\superscript{P}v_{n,p}+c_{n,p}\superscript{V}\left(1-v_{n,p}\right),
\end{equation}
where $c_{n,p}\superscript{P}$ and $c_{n,p}\superscript{V}$ are the cost to implement $p$ for $n$ physically and virtually, respectively. \revise{Their values in practical systems are determined by the specific hardware and software that are used by the MNO}{}. Generally, given an arbitrary fixed $i_{n,p}$,
\begin{equation}
	c_{n,p}\superscript{P}>c_{n,p}\superscript{V},\quad\forall (n,p)\in\mathcal{N}\times\mathcal{P}.
\end{equation}
The total isolation cost of slice $n$ is therefore,
\begin{equation}
	c_n=\sum\limits_{p\in\mathcal{P}}\left(c_{n,p}\superscript{ifr}+c_{n,p}\superscript{op}\right).
\end{equation} 

For every slice $n$, we define the quality of service $q_n$ and the security level $s_n$. We aim at minimizing the isolation cost:
\begin{mini!}[2]
	{\mathbf{I},\mathbf{T},\mathbf{V}}{\sum\limits_{n\in\mathcal{N}}c_n}{\label{prob:min_iso_cost}}{}
	\addConstraint{t_{n,p}+m_{n,p}=1,\forall (n,p)\in\mathcal{N}\times\mathcal{P}\label{con:shared_control}}
	\addConstraint{p_n\ge p_n\superscript{min},\forall n\in\mathcal{N}}
	\addConstraint{s_n\ge s_n\superscript{min},\forall n\in\mathcal{N},}
\end{mini!}
\revise{where $\mathbf{I}=[i_{1,1}, i_{1,2}\dots,i_{1,P}, i_{2,1}, i_{2,2}\dots,i_{2,P}\dots i_{N,P}]$ is the vector of isolation point selection, $\mathbf{T}=[t_{1,1}, t_{1,2}\dots,t_{1,P}, t_{2,1}, t_{2,2}\dots,t_{2,P}\dots t_{N,P}]$ the vector of control level specification, and $\mathbf{V}=[v_{1,1}, v_{1,2}\dots,v_{1,P}, v_{2,1}, v_{2,2}\dots,v_{2,P}\dots v_{N,P}]$ the vector of virtualization selection. Here, \eqref{con:shared_control} implies that the control over each protocol layer $p$ of a leased network slice $n$ is shared between the tenant and the MNO, and their controls are mutually exclusive to each other, which we have illustrated in Fig. 5.}{}

Note that let $i_1>i_2$, $m_1>m_2$, $t_1>t_2$, for all $(n,p)$ we have:
\begin{align}
	t_{n,p}\superscript{max}\vert_{i_{n,p}=i_1} &> t_{n,p}\superscript{max}\vert_{i_{n,p}=i_2}\label{eq:contro_level_bound}\\
	c_{n,p}\superscript{P}\vert_{i_{n,p}=i_1} &> c_{n,p}\superscript{P}\vert_{i_{n,p}=i_2}\label{eq:phy_inf_cost_dependency}\\
	c_{n,p}\superscript{V}\vert_{i_{n,p}=i_1} &> c_{n,p}\superscript{V}\vert_{i_{n,p}=i_2}\label{eq:virtual_inf_cost_dependency}\\
	c_{n,p}\superscript{op}\vert_{m_{n,p}=m_1} &> c_{n,p}\superscript{op}\vert_{m_{n,p}=m_2}\label{eq:op_cost_dependency}\\
	q_{n}\vert_{i_{n,p}=i_1} &> q_{n}\vert_{i_{n,p}=i_2}\label{eq:pfmc_iso_lvl}\\
	s_{n}\vert_{t_{n,p}=t_1} &> s_{n}\vert_{t_{n,p}=t_2}\label{eq:scrt_prtcl_stack}\\
	s_{n}\vert_{i_{n,p}=i_1} &> s_{n}\vert_{i_{n,p}=i_2}\label{eq:scrt_iso_lvl}
\end{align}
More specifically, \eqref{eq:contro_level_bound} \revise{implies that}{is implied} the constrains of isolation level and the upper bound of MNO control level. As a result, the more isolated, the more layers in the protocol stack can be securely controlled by the MNO. \eqref{eq:phy_inf_cost_dependency} and \eqref{eq:virtual_inf_cost_dependency} \revise{imply}{are implied} that the infrastructural cost of a network slice under certain level of protocol layer, no matter the protocol layer is physically or virtually implemented. Furthermore, the cost of network slice would increase along with the isolation. For example, it may cost more to maintain the protocol layer on an air-gap isolation independent server than to run it on a virtual machine. \eqref{eq:op_cost_dependency} \revise{shows that}{is showed} the operations cost in relating to the protocol layer which would increase along with the MNO control level. Since, it requires more effort in the VNF MANO module.
\eqref{eq:pfmc_iso_lvl} \revise{shows that}{is showed} the performance of a network slice that can be improved by raising the isolation level of its arbitrary protocol layer, since less loss will be caused by the resource scheduling among different slices sharing the same infrastructure under bear-metal or virtual machine.\eqref{eq:scrt_prtcl_stack} is referred to the fact that a network slice is more secure, when more of its control is granted to the tenant rather than the MNO. \eqref{eq:scrt_iso_lvl} is referred to the fact when a network slice is more secure, it is a better isolation from the other network slices.

\revise{It is worth to remark that $\mathbf{I}$, $\mathbf{T}$ and $\mathbf{V}$ are all defined on discrete sets, making the program \eqref{prob:min_iso_cost} non-convex and therefore rejecting conventional convex optimization problem solvers. Nevertheless, their domains are all finite, making it possible to solve \eqref{prob:min_iso_cost} with simple exhaustive search in cases where $N$, $P$ and $\vert\mathcal{T}\vert$ are small. For cases where the dimension is large and exhaustive search becomes computationally expensive, we can relax \eqref{prob:min_iso_cost} into a linear programming (LP) problem by extending the domains of $\mathbf{I}$, $\mathbf{T}$, $\mathbf{V}$ into continuous spaces through linear interpolation. Such linear programs are guaranteed to be efficiently solved with a polynomial time complexity. Thereafter, the optimal solution to the original problem \eqref{prob:min_iso_cost} can be obtained by rendering the optimum of relaxed LP, e.g. with the well-known branch-and-bound or cutting-plane algorithms.}{}

\subsection{Network Slice Planning Procedures}
\begin{figure}[!htbp]
	\centering
	\begin{subfigure}{.8\linewidth}
		\centering
		\includegraphics[width=\linewidth]{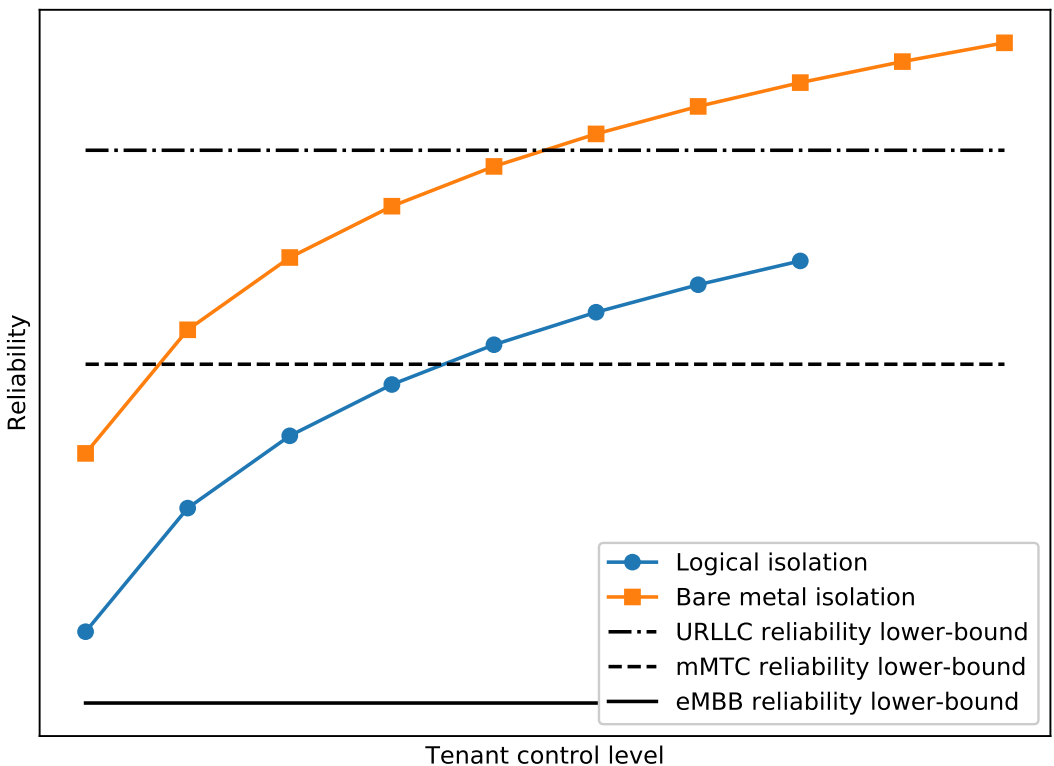}
		\caption{Checking the QoS satisfaction under different isolation plans}
		\label{fig:qos_check}
	\end{subfigure}\\
	\begin{subfigure}{.8\linewidth}
		\vspace{3mm}
		\centering
		\includegraphics[width=\linewidth]{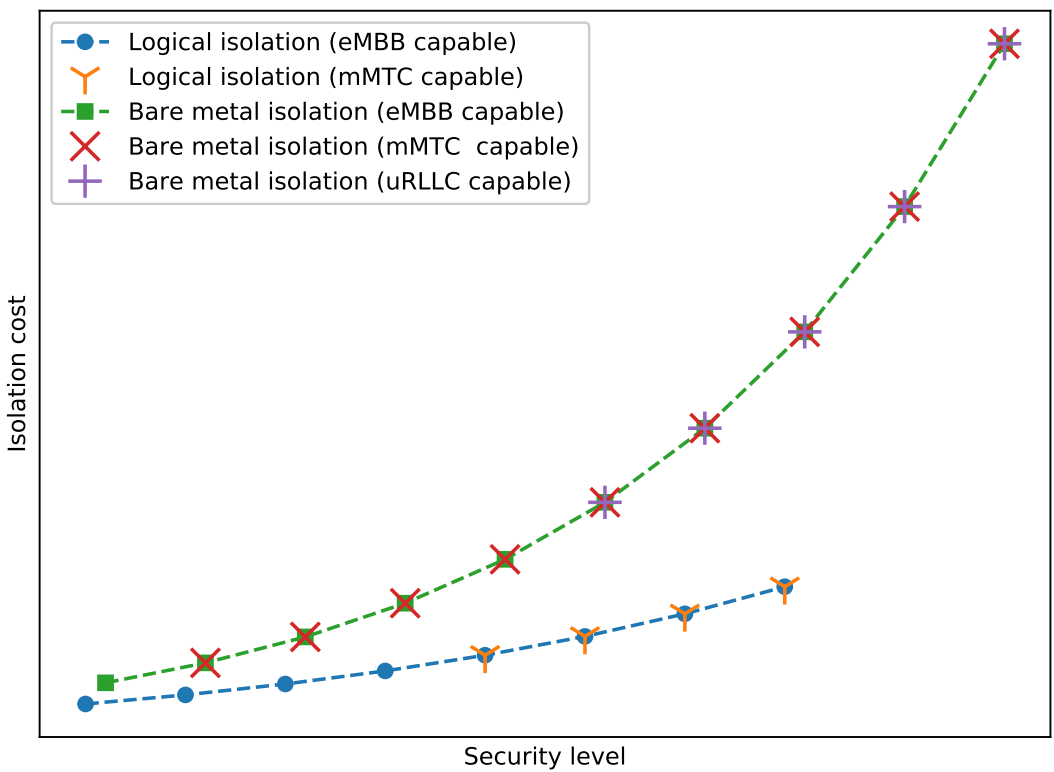}
		\caption{Selecting isolation plan w.r.t. the trade-off between security and cost}
		\label{fig:security_cost_tradeoff}
	\end{subfigure}
	\caption{How to select the appropriate NS isolation plan}
	\label{fig:iso_plan_selection}
\end{figure}

Network slice overview of deployment planning procedures can be referred in the following;
We begin with \revise{Figure}{figure}~\ref{fig:qos_check} gives the network slice isolation plan based on Quality of Service (QoS) satisfaction. Basically, the QoS satisfaction shall be conducted and aggregate individual QoS parameters. We take the reliability as an example which is shown in \revise{Figure}{figure}~\ref{fig:qos_check}. \revise{The}{There are} three horizontal \revise{lines}{axes} on the graph \revise{}{which} represent\revise{}{s} the minimal reliability requirements of a specific type of network slice service. \revise{More specifically}{For instance}, the level of service reliability \revise{requirement}{} of enhanced Mobile Broadband (eMBB) network slice is \revise{}{represented on} the lowest \revise{among the three, because eMBB}{horizontal line that} may accept high latency network QoS and a certain level of packet loss. The level of service reliability of massive Machine-Type Communications (mMTC) network slice is \revise{on a}{represented in the} mid-level\revise{, since mMTC deploys}{. It has} a large amount of devices connections with short messages transmission and with no re-transmission policy characteristics\revise{, since the duty cycle of mMTC devices}{. It is because of the duty cycle} may be very short. \revise{These features of mMTC are asking for a}{Therefore, we can categorize mMTC network slice that requires} better service reliability \revise{than that of the}{} eMBB network slice. Last but not the least, the highest \revise{}{requirement of} service reliability is \revise{}{required by} Ultra Reliable Low Latency Communications (URLLC) network slice\revise{, which}{ is represented in the highest-level. URLLC } very often applies to deliver critical infrastructure \revise{while commonly providing}{which also requires to provide } certain level of defense mechanisms.  
Figure~\ref{fig:security_cost_tradeoff} gives an other network slice deployment consideration. We often \revise{see}{seen} the network slice deployment using air-gap isolation, logical isolation, combination of isolation methods. Obviously, each network slice would have a set of contains with different protocol stack that can be controlled by tenant or MNO. From the MNO point of view, when tenant having more control of the network slice which is better to apply air-gap isolation due to having a better reliability than logical isolation. Figure~\ref{fig:security_cost_tradeoff} provides a result in comparing with air-gap (bear-metal) isolation and logical isolation. Therefore, in terms of considering from isolation, when we have the same isolation level and the reliability would increase with the tenant control level. On the other hand, when a network slice has the maximal tenant control level that requires the isolation level to be maximal as well. It can use air-gap isolation that is shown in the \revise{Figure}{figure}~\ref{fig:security_cost_tradeoff}. 

According to the results, obviously, the tenant can remove some of the isolation options. For example, in general, logical isolation would not be possible on deploying to URLLC services due to performance requirement, the only possible solution would be applied the air-gap isolation with physical resources. We can further to discuss the trad-off which is shown in \revise{Figure}{figure}~\ref{fig:security_cost_tradeoff}. Basically, the tenant can chooses from all available options with regard to its cost of security and defense preference. Typically, tenant may demand security hardening network slice, but the cost of isolation would be directly proportional to security hardening which is shown in \revise{Figure}{figure}~\ref{fig:qos_check}. Figure~\ref{fig:qos_check} is also indicated the upper bound of the security level in relating to the upper bound of tenant control level, i.e. by the isolation level - bare metal isolation can achieve more than logical isolation. In order to achieve the better security level, it is always use air-gap (bare-metal) isolation than apply logical isolation. However, it may cost most and less flexibility.

\section{Conclusion \revise{and Outlooks}{}}
\label{con}
When deploying NSaaS, MNO must resolve various levels of complex deployment and operation issues in order to provide a secure service to the vertical industries (tenants). Although, 3GPP has thoroughly laid out the 5G standalone architecture and provides network slice enablement functions, it has not yet identified the common practice and the security by design in operating the NSaaS. With its main functions based on virtualization and containerization technologies, NSaaS gives flexibilities and agilities to the telecommunication infrastructure; however, it also introduces meanwhile a number of new risk factors and widens the attack surface. In this paper, we have explored and addressed the complexity and challenges of risk factors and the attack surface in eight layers of NSaaS, which allow MNOs and tenants to identify the defense mechanisms on each of the particular network slice. Especially, for a NSaaS platform in operation, each of the network slice should put certain level of isolation in reflecting the level of protocol stack control from either the MNO or the tenant. Also, these deployed isolation methods are related to the overall protection of the network infrastructure, and defense mechanisms shall be embedded with the network architecture, e.g., micro-segmentation etc. We have also developed a mathematical model to represent the relationship between isolation level and the control distribution over MNO and tenant. This model can be used to guide the MNO and tenant in designing the SLA regarding their control levels and the isolation cost of the deployed network slice. The result shows that air-gap isolation provides the ideal performance of network slice deployment, but it also has the highest cost due to the under-utilization of resources. \revise{As a possible research direction for our future work, it is of our great interest to evolve the qualitative models we have developed in this article into quantitative ones. To do so, case studies upon specific deployment scenario and practical applications must be carried out.}{}

\section*{Abbreviations}
{The following abbreviations are used in this manuscript:\\
	\noindent 
	\begin{tabular}{@{}ll}
		3GPP & Third Generation Partnership Project \\
		AD & Administrative Domain \\
		eMBB & enhanced Mobile Broadband \\
		GSMA & GSM Association \\
		ITU & International Telecommunication Union \\
		mMTC & Massive Machine-Type Communications\\
		MNO & Mobile Network Operator\\
		NFV & Network Function Virtualization \\
		NSaaS & Network Slice as a Service \\
		QoS & Quality of Service \\
		SDN & Software-Defined Network \\
		SLA & Service Level Agreement \\
		VNF & Virtual Network Function \\
		URLLC & Ultra Reliable Low Latency Communications
\end{tabular}}

%
\vfill

\begin{IEEEbiography}[{\includegraphics[width=1in,height=1.25in,clip,keepaspectratio]{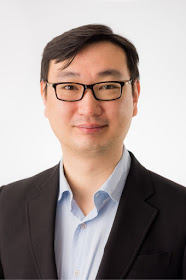}}]{Stan Wong} joined Hong Kong Telecom Limited in 2019 as Assistant Vice President and responsible for PCCW Group Technology Innovations, 5G security and deployment. He received his Ph.D. from University of London, King’s College London (KCL) in 2009. He was leading the European Commission H2020 5G Novel Radio Multiservice Adaptive Network Architecture (NORMA) as a security leader, and 5G Infrastructure Public Private Partnership (PPP) security working group lead Trust chapter editor. He authored the 5G Trust Model in NGMN. Currently, he represents HKT in various standardization development organization (SDO) such as GSMA, NGMN, ETSI, 3GPP SA3 and Wireless Alliance.
\end{IEEEbiography}


\begin{IEEEbiography}[{\includegraphics[width=1in,height=1.25in,clip,keepaspectratio]{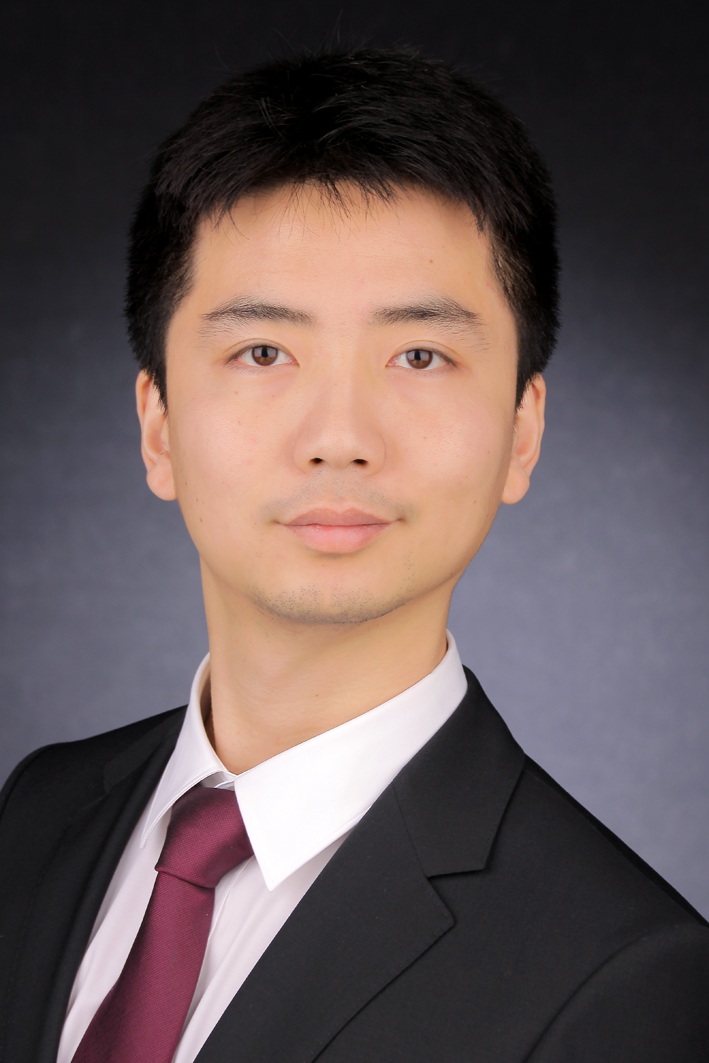}}]{Bin Han} received in 2009 his B.E. degree from Shanghai Jiao Tong University, M.Sc. in 2012 from Technical University of Darmstadt, and in 2016 the Ph.D. degree from Karlsruhe Institute of Technology. Since July 2016 he has been with University of Kaiserslautern as Postdoctoral Research Fellow and Senior Lecturer. His research interests are in the broad area of wireless communication and networking, with a special current focus on the topics of network slicing, timely and reliable communication, and mobile edge computing. He is the author of over 40 research papers and book chapters, and has participated in multiple EU collaborative research projects. He serves as a Guest Editor of Network (MDPI) and Electronics (MDPI), the TPC Chair of the Workshop on Next Generation Networks and Applications, and a TPC Member of GLOBECOM, EuCNC, and European Wireless. He is a Senior Member of IEEE, a Member of the IEEE Communication Society, and a Member of the IEEE Information Theory Society.
\end{IEEEbiography}


\begin{IEEEbiography}[{\includegraphics[width=1in,height=1.25in,clip,keepaspectratio]{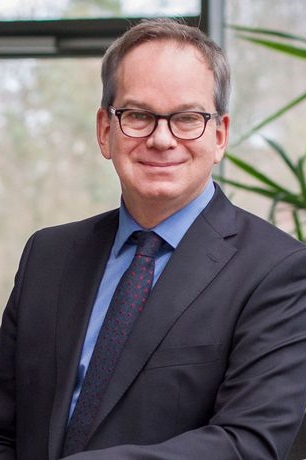}}]{Hans D. Schotten} received the Ph.D. degree from the RWTH Aachen University of Technology, Germany, in 1997. From 1999 to 2003, he worked with Ericsson. From 2003 to 2007, he worked with Qualcomm. He became the Manager of a R\&D Group, a Research Coordinator for Qualcomm Europe, and the Director for Technical Standards. In 2007, he accepted the offer to become the Full Professor with the University of Kaiserslautern. In 2012, he became a Scientific Director of the German Research Center for Artificial Intelligence (DFKI) and the Head of the Department for Intelligent Networks. He served as the Dean of the Department of Electrical Engineering, University of Kaiserslautern from 2013 until 2017. He has authored more than 200 papers and participated over 40 European and national collaborative research projects. Since 2018, he has been the Chairman of the German Society for Information Technology and a Member of the Supervisory Board of the VDE.
\end{IEEEbiography}



\begin{thebibliography}{999}
	
	\bibitem{EURSIP:NS5G}
	Prashant Subedi, Abeer Alsadoon, P. W. C. Prasad, Sabih Rehman, Nabil GiweliMuhammad Imran and Samrah Arif, ``Network Slicing: a next generation 5G perspective'', in EURASIP Journal on Wireless Communications and Networking, Vol. 2021 Issue 1, pp. 1-26, 2021.
	
	\bibitem{Wiley:NSSecC}
	Vitor A. Cunha, Eduardo da Silva, Marcio B. de Carvalho, Daniel Corujo, Joao P. Barraca, Diogo Gomes, Lisandro Z. Granville, and Rui L. Aguiar, ``Network slicing security: Challenges and directions,'' in Wiley Internet Technology Letters, Vol. 2, Issue 5, September/October 2019.
	
	\bibitem{IEEE:5GNSSec}
	Ruxandra F. Olimid and Gianfranco Nencioni, ``5G network slicing: A security overview,'' IEEE Access, Vol. 8, pp. 99999-100009, May 2020.
	
	\bibitem{MDPI:NSC}
	Jose Ordonez-Lucena, Pablo Ameigeiras, Luis M. Contreras, Jes\`us Folgueira, and Diego R. L\'opez, ``On the rollout of network slicing in carrier networks: A technology radar,'' in MDPI Sensors, Vol. 21, Issue 23, 2021. 
	
	\bibitem{JNSM:5GMD}
	Jos\'e Mar\'ia Jorquera Valero, Pedro Miguel S\'anchez S\'anchez, Alexios Lekidis, Javier Fernandez Hidalgo, Manuel Gil P\'erez, M. Shuaib Siddiqui, Alberto Huertas Celdrán, and Gregorio Martínez P\'erez,``Design of a security and trust framework for 5G multi-domain scenarios,'' in Journal of Network and Systems Management, Vol. 30, Article number: 7 (2022), October 2021.
	
	\bibitem{MDPI:NSSecCrlt}
	Tomasz Wichary, Jordi Mongay Batalla, Constandinos X. Mavromoustakis, Jerzy Zurek, and George Mastorakis, ``Network slicing security controls and assurance for verticals,'' in MDPI Electronics, Volume 11, Issue 2, 2022.
	
	\bibitem{IEEE:CAuC5GNS}
	Chun-I Fan, Yu-Tse Shih, Jheng-Jia Huang, and Wan-Ru Chiu, ''Cross-network-slice authentication scheme for the 5th Generation mobile communication system,'' in IEEE Transactions on Network and Service Management, Vol. 18, No. 1, pp. 701-712, March 2021.
	
	{\color{\highlightcolor}
		\bibitem{CESAR:NSSecurityAI}
		Luis Suárez, David Espes, Philippe Le Parc, Fr\'ed\'eric Cuppens, Philippe Bertin, and Cao Thanh Phan, ``Enhancing network slice security via Artificial Intelligence: Challenges and solutions,'' in C\&ESAR 2018 Conference, Rennes, France, November 2018.
	}
	
	\bibitem{Wiley:SS5GNS}
	Ali Esmaeily and Katina Kralevska, ``Small-scale 5G testbeds for network slicing deployment: A systematic review,'' in Wiley Wireless Communications and Mobile Computing, Vol. 2021, pp. 1-26, 2021.
	
	\bibitem{GSMA:NSintro}
	GSM Association, Whitepaper, ``An Introduction to Network Slicing'', 2017.
	
	\bibitem{GSMA:NSGT}
	GSM Association, NG.116, ``Generic Network Slice Template v6.0'', 2021.
	
	\bibitem{IEEE:CloudNet}
	Polychronis Valsamas, Panagiotis Papadimitriou, Ilias Sakellariou, Sophia Petridou, Lefteris Mamatas, Stuart Clayman, Francesco Tusa, and Alex Galis, ``Multi-PoP network slice deployment: A feasibility study'', in IEEE 8th International Conference on Cloud Networking (CloudNet), November 2019.
	
	
	\bibitem{CIS:url}
	https://www.cisecurity.org/resources/?type=benchmark
	
	\bibitem{NIST:SC}
	NIST, Cybersecurity and Infrastructure Security Agency, ``Defending Against Software Supply Chain Attacks'', April 2021.
	
	
	\bibitem{GSMA:SC}
	GSM Association, Whitepaper, ``Telecommunication Supply Chain Status Update'', September 2020.
	
	\bibitem{NCSC:urlSC}
	https://www.ncsc.gov.uk/collection/supply-chain-security
	
	\bibitem{EU:50600}
	BS EN 50600-1:2019. ``Information technology -- Data centre facilities and infrastructures General concepts''.
	
	\bibitem{ISO:22237}
	ISO/IEC 22237-1:2021, ``Information Technology -- Data centre facilities and infrastructure - Part I General Concept''.
	
	{\color{\highlightcolor}
		\bibitem{IEEE:IsoApp}
		Chunyang Ye, S.C. Cheung, and W.K. Chan, “Sifter: A Service Isolation Strategy for Internet 
		Applications”, in IEEE TRANSACTIONS ON SERVICES COMPUTING, VOL. 14, NO. 5, pp: 1545 - 1557, Sept.-Oct. 2021,
		
		\bibitem{IEEE:know} 
		Sahand Hariri, Matias Carrasco Kind, and Robert J. Brunner, “Extended Isolation Forest”, 
		IEEE TRANSACTIONS ON KNOWLEDGE AND DATA ENGINEERING, VOL. 33, NO. 4, pp: 1479 -1489, APRIL 2021.
		
		\bibitem{NASA:ML} 
		James R. Carnes, Douglas H. Fisher, “Machine Learning Techniques for Fault Isolation and Sensor Placement”, NASA Technical Reports Server (NTRS), November 16 1992.}
	
	\bibitem{Wiley:e2eNS}
	Sabra Ben Saad, Adlen Ksentini, and Bouziane Brik, ``An end-to-end trusted architecture for network slicing in 5G and beyond networks,'' in Wiley Security and Privacy, Vol. 5, Issue 1, January/February 2022.
	
	\bibitem{3GPP:TS38300}
	3GPP, TS 38.300, ``NR; NR and NG-RAN Overall description; Stage-2'', Release 16.
	
	\bibitem{3GPP:TS33501}
	3GPP, TS 33.501, ``Security architecture and procedures for 5G System'', Release 16.
	
\end{thebibliography}
\end{document}